\documentclass[12pt]{article}

\usepackage{epsfig}

\newcount\xrefpos \xrefpos=0
\newcount\yrefpos \yrefpos=0
\newcount\xput \xput=0
\newcount\yput \yput=0

\def\refpos#1 #2 #3{\global\xrefpos=#1 \global\yrefpos=#2
                         \rlap{$\smash{#3}$}}
\def\put #1 #2 #3{\xput=#1 \yput=#2
                  \advance\xput by -\xrefpos
                  \advance\yput by -\yrefpos
                  \rlap{\kern\the\xput truebp
                        \vbox to 0pt{\vss\hbox{$\displaystyle #3$}
                        \kern\the\yput truebp}}}
\def\beginlabels\refpos#1\endlabels{\hbox{$\refpos#1$}}

\newcommand{\lb}{\langle}
\newcommand{\rb}{\rangle}

\renewcommand{\d}{\ensuremath{{\rm d}}}

\newcommand{\ra}{\ensuremath{\rightarrow}}

\newcommand{\be}{\begin{equation}}
\newcommand{\ee}{\end{equation}}
\newcommand{\beq}{\begin{equation*}}
\newcommand{\eeq}{\end{equation}}
\newcommand{\ba}{\begin{eqnarray}}
\newcommand{\ea}{\end{eqnarray}}

\begin{document}

\begin{titlepage}

\bigskip
\hskip 4.8in\vbox{\baselineskip12pt \hbox{hep-th/yymmnnn}}

\bigskip
\bigskip
\bigskip

\begin{center}

{\Large \bf  On Bubbles of Nothing in AdS/CFT}
\end{center}

\bigskip
\bigskip
\bigskip

\centerline{\bf Jianyang He and Moshe Rozali }

\bigskip
\bigskip
\bigskip

\centerline{ \it Department of Physics and Astronomy}
\centerline{\it University of British Columbia} \centerline{\it
Vancouver, British Columbia V6T 1Z1, Canada}

\bigskip
\bigskip

\begin{abstract}
\vskip 2pt We discuss non-perturbative instabilities, mediated by
bubbles of nothing, in the context of the AdS/CFT correspondence. By
exploring the phase diagram of such decays we give an interpretation
of the process in terms of an effective potential in a quantum
mechanics of a single variable, a winding mode condensate. The decay
corresponds to a conventional first order transition, the true
vacuum being a non-geometrical "nothing" state.
\end{abstract}

\end{titlepage}


\baselineskip=18pt \setcounter{footnote}{0}

\section{Introduction and Motivation}

The AdS/CFT correspondence (for a review see \cite{ads}) provides a
non-perturbative background independent definition of quantum
gravity in asymptotically AdS spaces. It is useful then to
investigate quantum gravity issues in this context. In this paper we
examine non-perturbative instabilities mediated by bubbles of
nothing \cite{witten,vijay1}. Semiclassical analysis suggests these
correspond to vacuum decay, and indeed the false vacuum was
identified in \cite{vijay2} as the topological black hole
\cite{top}. It is interesting then to discuss the process, and its
place in the full non-perturbative definition of the theory. One
goal is to clarify the role of the semi-classical analysis in the
full theory, as such non-perturbative instabilities may play a
crucial role in flux compactifications or in the context of eternal
inflation.

In our context, the dual field theory is a conventional
(supersymmetric, conformal) field theory formulated in curved space,
namely $dS_3 \times S^1$ \cite{vijay2}. This suggests that long
distance features of the theory are captured in the matrix quantum
mechanics of the lowest lying mode, which is the Wilson line
wrapping the circle, which we call $U$. The time in that quantum
mechanics is the non-compact time direction of the deSitter
space\footnote{ In particular we do not expect the features of the
decay to be captured by matrix integrals,  such as the ones in
\cite{mark1}, appropriately analytically continued, as suggested for
example in \cite{vijay2,them}. The process of bubble nucleation is
time dependent, therefore degrees of freedom along the time
direction of $dS_3$ should not  be integrated out. More on that in
\cite{us}.}. A priori that quantum mechanics has a time dependent
Hamiltonian, but we will find that for our purposes that time
dependence plays only a relatively minor role.

In a companion paper \cite{us} we will calculate the effective
action of that matrix quantum mechanics in the weak 'tHooft
coupling, small volume limit, following the work in \cite{mark1}.
Our purpose in this paper is to collect evidence that the reduction
to a quantum mechanics of a single degree of freedom $u$ (defined as
the condensate of the $Tr(U)$) captures the essential features of
the process, even in the supergravity (large 'tHooft coupling)
limit.

To this end we examine the phase diagram of the theory as function
of the circle's radius and three R-charge chemical
potentials\footnote {For previous analysis of the phase diagram see
\cite{Biswas}, for discussion of charged bubbles see also
\cite{dim}.}. We find that generically the bulk analysis indicates a
first order transition. The transition becomes enhanced (the bounce
action becomes small) at specific loci in the phase diagram; we find
that in precisely those loci the bulk developed a winding tachyon,
which is precisely the bulk mode dual to $u$. These loci then
correspond to the disappearance of the barrier in the effective
potential, causing the instability to become perturbative and driven
by the process of (winding) tachyon condensation. This is precisely
what one would expect in the quantum mechanical model of the winding
condensate $u$.

This analysis can be taken as an evidence for the conjecture in
\cite{Horowitz:2005vp,Horowitz:2006mr} that the end point of the
tachyon condensation is indeed the bubble of nothing. Inside that
bubble is the true vacuum, the "nothing state", which is then a
winding mode condensate. That state is inherently stringy, and does
not posses a conventional spacetime interpretation. This is despite
the fact that no strong curvatures exist in the bubble spacetime. It
would be interesting to further discuss this non-geometrical phase
in terms of its field theory dual, understanding for example the
disappearance of spacetime in this state.

The outline of this note is as follows. In the next section we
discuss the general bubble with up to three chemical potentials
turned on. We demonstrate the features of the phase diagram for a
few specific cases, and discuss the general qualitative features.
Finally, we discuss the explanation of all these features in terms
of the quantum mechanics of the variable $u$ and its purported
effective potential. We find that all those qualitative features,
such as the phase boundaries, can be explained by a conventional
quantum mechanical model, where the process is simply the familiar
decay of the false vacuum. We present our conclusions and directions
for future research in the final section.

\section{Bubbles of Nothing in AdS/CFT}

In this section we review and extend the bulk analysis of the decays
mediated by the various bubble of nothing (BON) solutions. We focus
on the qualitative features of the relevant gauge theory - in all
cases the maximally supersymmetric Yang-Mills theory, formulated in
curved space, in the large $N$ and strong 'tHooft coupling limits.
Those qualitative features will be compared to those in the  weakly
coupled gauge theory in \cite{us}.

We will concentrate on exploring the phase diagram, looking for the
regions in parameter space where the instability exists, and
interpret these qualitative features in terms of a candidate
effective potential, such as the one existing in the dual gauge
theory. The non-perturbative instability entails an existence of an
appropriate bounce solution, a Euclidean solution of the equations
of motion with a single (non-conformal) negative mode. The negative
mode signals the existence of a local maximum in the effective
potential. Additionally we compare the Euclidean action of the
bounce with that of the false vacuum: in order for the decay rate is
small, corresponding to a meta-stable false vacuum, the action of
the bounce has to be higher than that of the false vacuum. When the
actions become comparable the false vacuum is no longer meta-stable.

\subsection{R-Charged Bubbles}

Let us start with the most general black hole metric carrying up to
3 unequal R-charges. The bubble solutions we are interested in are
formed by double analytic continuation as described below. The
Lorentzian black hole metric is \cite{Rcharges}\be
\label{eq:R_metric} \d s^2 = -(H_1 H_2 H_3)^{-2/3}f \d t^2 +(H_1 H_2
H_3)^{1/3}\left(f^{-1}\d r^2+r^2\d\Omega_{3,k}\right),\ee where \be
f=k-\frac{\mu}{r^2}+\frac{r^2}{l^2}  H_1 H_2 H_3,~~~~~~
H_i=1+\frac{q_i}{r^2}, ~~~i=(1,2,3), \ee

Here $q_i$ are 3 charge parameters, related to the physical charges
in a manner specified below. To obtain a bubble of nothing solution
 we perform double analytic continuation. The reality conditions on all fields distinguish
 between the bubble
interpretation and the more familiar black hole (thermal) one, which
was analyzed in \cite{Rcharges}, thus resulting in a different phase
diagram. In our case the circle parameterized by $\chi=it$ is
interpreted as a spatial direction of the geometry, rather than
Euclidean time, therefore the gauge field component in that
direction $A_\chi$ has to be real. To ensure that we take the charge
parameters $q_i$ to be negative, as opposed to having them take
positive values for the black holes analyzed in \cite{Rcharges}.

 The Euclidean metric approaches
asymptotically $M_k \times S^1$, where $S^1$ is a circle of
asymptotic radius $\beta$ (which is a parameter of the solution),
and $M_k$ is an homogeneous space of constant curvature $k$ and
metric $\d \Omega_{3,k}$. For the case $k=1$ it is $S^3$, if $k=0$
it is $R^3$, and if $k=-1$ it is a hyperbolic space $H_3$. Upon
analytic continuation, one of the coordinate of $M_k$ becomes
timelike, and the solution resemble a bubble, with the $S^1$ being
interpreted as a spatial circle. In the following we shall
concentrate on the solutions with $k=1$, in which case the
analytically continued metric on $M_k$ is that of $dS_3$, the 3
-dimensional deSitter space.

 In
addition to the metric the solution has three scalar fields $X^i$
and gauge fields $A^i_\mu$, which  are of the form \be
X^i=H_i^{-1}(H_1 H_2 H_3)^{1/3}, ~~~ A^i_t=\frac{\tilde
q_i}{r^2+q_i}+\phi_i, ~~~i=1,2,3. \ee The constants $\phi_i$ are
adjusted such that the gauge potential at the Euclidean origin is
zero. In that case they equal the gauge potentials at infinity, and
thus are parameters of the dual gauge theory\footnote{No such
parameters, or charges, are associated with the scalar fields
$X_i$.}.

The physical charges  $\tilde q_i$ are given in terms of the
parameters $q_i$ as \be
\label{pot1}\tilde{q}^2_i=q_i(r_+^2+q_i)\left[
1+\frac{1}{r_+^2}\prod_{j\neq i}(r_+^2+q_j)\right],\ee where $r_+$
is the location of the outer horizon, namely the largest root of
$f(r)$ defined above. We will mostly work in "grand-canonical
ensemble" where the fixed quantities are the potentials at infinity,
given by \be \label{pot2}\phi_i \equiv - A_t^i(r_+)=-
\frac{\tilde{q}_i}{r_+^2+q_i}.\ee Using the notations of
\cite{Rcharges} we take $\phi_i$ to be purely imaginary, which
ensures the correct reality conditions upon analytic continuation to
the Lorentzian bubble spacetime.

To identify the false vacuum, decaying via the Euclidean solution,
interpreted as a bounce, we look at the asymptotic behavior of the
fields. The metric behaves asymptotically the same as in the
solutions in \cite{vijay2}, therefore in all cases the spacetime
decaying is the topological black hole which has a non-contractible
circle in the geometry (it is obtained from AdS by appropriate
identifications, as discussed in \cite{vijay2}). The scalars fall
off rapidly at infinity, and the gauge fields approach a constant.
We conclude therefore that the false vacuum is the topological black
hole with constant gauge potentials around the non-contractible
circle\footnote{Furthermore the variables $|\phi_i|$ are periodic in
units of $\beta^{-1}$.}.

To map out the general phase diagram we are first interested in the
region of parameter space for which the Euclidean solution exists
and has a single non-conformal negative mode. In all cases we fix
the asymptotic radius $\beta$ and the value of the potentials at
infinity $\phi_i$.

In our coordinates there is a possible conical singularity at
$r=r_+$. Demanding regularity at the Euclidean origin determines as
usual the asymptotic radius $\beta$. In this case we find \be
\beta=2\pi \frac{r_+^2\sqrt{\prod_i(r_+^2+q_i)}}{2r_+^6+(1 +\sum_i
q_i)r_+^4-\prod_i q_i},\ee where we set for  convenience  the AdS
radius to unity, $l=1$. Together with the formulas (\ref{pot1},
\ref{pot2}) this determines the region of parameter space (spanned
by $\beta$ and $\phi_i$ for which a candidate Euclidean solution
exists).

To confirm that a candidate Euclidean solution is indeed a bounce
one needs to check the existence of a non-conformal negative mode.
In general this is a difficult problem, solved for the uncharged
case in the classic reference \cite{gpy}. However, it is known that
"thermodynamic instability" means that the solution is a local
maximum of the free energy,  and therefore it is a sufficient
condition for the existence of a negative mode \cite{reall}.
Therefore,  in order to check for the possibility of a negative mode
we check the Hessian of the Euclidean action (with respect to the
field theory parameters) in the region of parameter space relevant
for the bubble interpretation.

Finally, to check for the existence of a barrier in field space we
calculate
 the Euclidean action (relative to the false vacuum
 which is locally simply AdS space). The Euclidean
action then can be calculated directly from the  above solution, it
is found to be \cite{Rcharges}  \be
I=\beta(M-\sum^3_{i=1}\tilde{q}_i \phi_i)-S,\ee where the mass $M$
is given by \be M=\frac{3}{2}\mu+\sum q_i=\frac{3}{2}(r_+^2+r_+^4
\prod ^3_{i=1} H_i)+\sum^3_{i=1}q_i,\ee and the
"entropy"\footnote{This quantity has no interpretation as entropy in
the analytic continuation leading to the bubble solution.} $S$ is
\be S=\frac{A}{4G_N}=2\pi\sqrt{\prod (r_+^2+q_i)}.\ee

\subsection{Uncharged Case}

The simplest case of the uncharged bubble was discussed extensively
in \cite{vijay1,vijay2}, and we discuss it here briefly. In that
case the Euclidean metric is \be \d s^2=f\d \chi^2+f^{-1}\d
r^2+r^2(\d\theta^2+\sin^2\theta\d\Omega_2), \ee where \be
f=1-\frac{\mu}{r^2}+r^2.\ee The phase diagram is one dimensional,
characterized by the size of the asymptotic circle $\beta$, which is
given by \be \beta=\frac{2\pi r_+}{1+2r_+^2} \ee where the horizon
location $r_+$ can take any positive value. We see then that since
this function attains a maximum at $\beta_{crit} =\pi/\sqrt{2}$,
there cannot be a non-perturbative instability for
$\beta>\beta_{crit}$. For $\beta<\beta_{crit}$ one has two possible
solutions (two values of $r_+$), and one has to check for the
existence of negative mode.

 The Euclidean action as a function of the parameter $r_+$ is given by \cite{vijay1}  \be
I =-\frac{\pi r_+^3(r_+^2-1)}{1+2r_+^2},\ee where we have omitted an
irrelevant reference constant. As explained above, to check for the
existence of a negative mode we calculate the Hessian of the action,
in this case this is simply the second derivative at fixed
$\beta$\footnote{This quantity  would be the fixed temperature
specific heat in the thermal interpretation.}\be \frac{\partial ^2
I_{Euc}}{\partial (r_+^2)^2}\left| _{\beta}=\frac{3\pi
(2r_+^2-1)}{2r_+(1+2r_+^2)}\right.
\ee We see that for $\beta<\beta_{crit}$ there are two possible
solutions, and one of them  has a negative mode, corresponding to an
instability. Therefore we conclude that for every
$\beta<\beta_{crit}$ a bounce exists.

Additionally, as  showed in figure 1, the Euclidean bubble has
positive action (relative to the false vacuum, that is the
topological black hole) for  $r_+<1$, which includes all the
relevant region of instability. The action becomes small,
corresponding to the disappearance of the barrier, for small radii
$\beta$, signalling the onset of tachyon condensation as an
alternative mode of instability, as discussed below.

\begin{figure}[h]
         \beginlabels\refpos 0 0 {}
                     \put 320 -90 {r_+^2}
                     \put 130 -40 {I_{bubble}}
         \endlabels
         \centerline{
         \psfig{figure=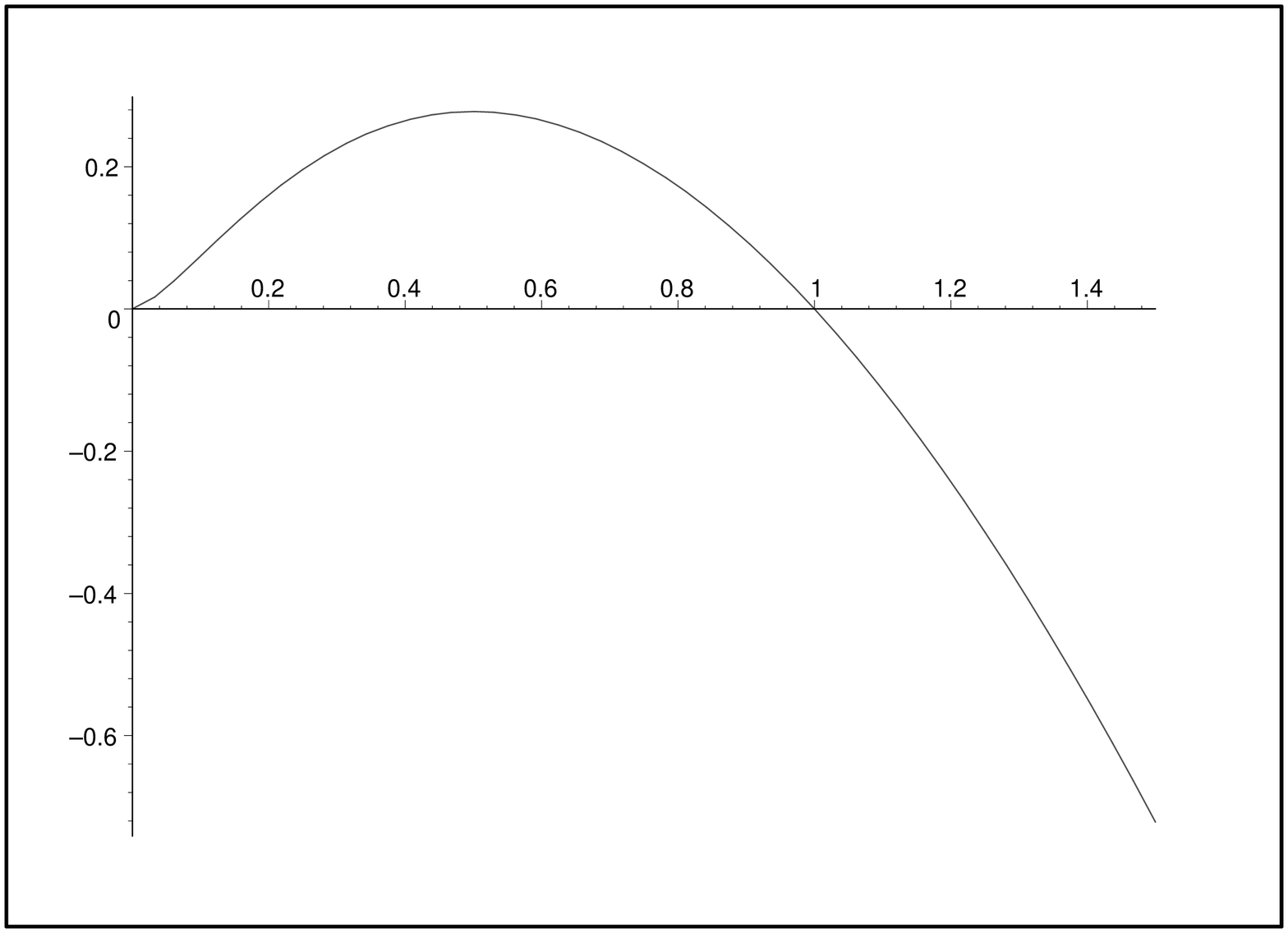,height=3.5in,angle=-90}}
         \vskip0.5cm
         {\footnotesize Figure 1.  The Euclidean action of the bubbles of size $r_+$. When the action
         is positive the false vacuum is metastable. The conditions for negative Euclidean mode are
         satisfied for $r_+^2 < \frac{1}{2}$. }
\end{figure}

To make connection with the conjecture in \cite{Horowitz:2005vp},
let us discuss the range of parameters for which a winding tachyon
appears in the geometry. Since the analysis depends only on the
geometry, and not on the background fields, it applies to all cases
below as well.

We start with the false vacuum, namely the topological black hole
\cite{vijay2} \be \d s^2= (1+\frac{r^2}{l^2})^{-1} \d
r^2+\beta^2(1+\frac{r^2}{l^2})\d\chi^2 + r^2 \d\Omega_3, \ee where
$l$ is the  AdS radius,  $\beta$ is the periodicity of the circle,
so that $\chi \sim \chi+2\pi$. The size of the circle is \be
L(r)=\beta \sqrt{1+\frac{r^2}{l^2}}. \ee It goes to a constant  at
$r\ra 0$, and to infinity at the boundary.

For a winding  tachyon to condense, two conditions are needed. The
size of the circle  has to be  of the string scale $l_s$, and it has
to vary slowly (compared to the string scale). These conditions (for
$l \gg l_s$ ) are satisfied in the regime $\beta<l_s$. This is
precisely the locus of the phase diagram for which the Euclidean
action becomes small, signalling the disappearance of the barrier in
field space\footnote{The analysis is consistent with the winding
tachyon found to exist in \cite{Horowitz:2006mr}, since the AdS
soliton is the limit for which the $dS_3$ radius of curvature
diverges, or by conformal transformation, the limit where $\beta \ra
0$. For additional support for the conjecture in
\cite{Horowitz:2005vp} see also \cite{tak}.}.

\subsection{One Charge Case}

Now let turn on a single charge, say $q_1=q$, and $q_2=q_3=0$, the
field theory parameters are given by \ba \tilde{q}^2 &=&
q(r_+^2+q)(1+r_+^2), \nonumber \\\phi&=&-\frac{\tilde{q}}{r_+^2+q},
\nonumber \\ \beta&=&\frac{2\pi\sqrt{r_+^2+q}}{2r_+^2+1+q}\ea

The bubble solution, obtained by double analytic continuation from
the black hole $t\ra i\chi$, $\theta\ra\frac{\pi}{2}+i\tau$, is
given by \cite{Biswas} \be \d s^2=H^{-2/3}f\d
\chi^2+H^{1/3}(f^{-1}\d
r^2-r^2\d\tau^2+r^2\cosh^2\tau\d\Omega_2),\ee where \be
H=1+q/r^2,\hspace{1cm} f=1-\frac{\mu}{r^2}+r^2 H.\ee  and $q<0$ in
the bubble interpretation. And it is easy to get the action and
Hessian \ba \label{eq:I_onecharge_bubble}
I&=&\beta(M-\tilde{q}\phi)-S=-\frac{\pi
R\sqrt{R+q}(R-1+q)}{2R+1+q}, \nonumber \\
Hessian&=&\frac{\pi^2(3R^2+3(1+q)R+4q)(2R^2+(1+q)R-(1-q)^2)}{4q(2R+1+q)^2(R+q)(1+R)}.\ea
where we defined $R= r_+^2$.

First we look at the existence of bubbles in the parameter space
spanned by $\beta$ and $\beta |\phi|$. This is the shaded region in
figure 2. We see that, like in the neutral case, the bubbles exist
for a finite region in parameters space, and in particular there is
always a maximum radius $\beta_{max}$ for which the instability
disappears.

\begin{figure}[h]
         \beginlabels\refpos 0 0 {}
                     \put 330 -190 {\beta}
                     \put 120 -40 {\beta|\phi|}
         \endlabels
         \centerline{
         \psfig{figure=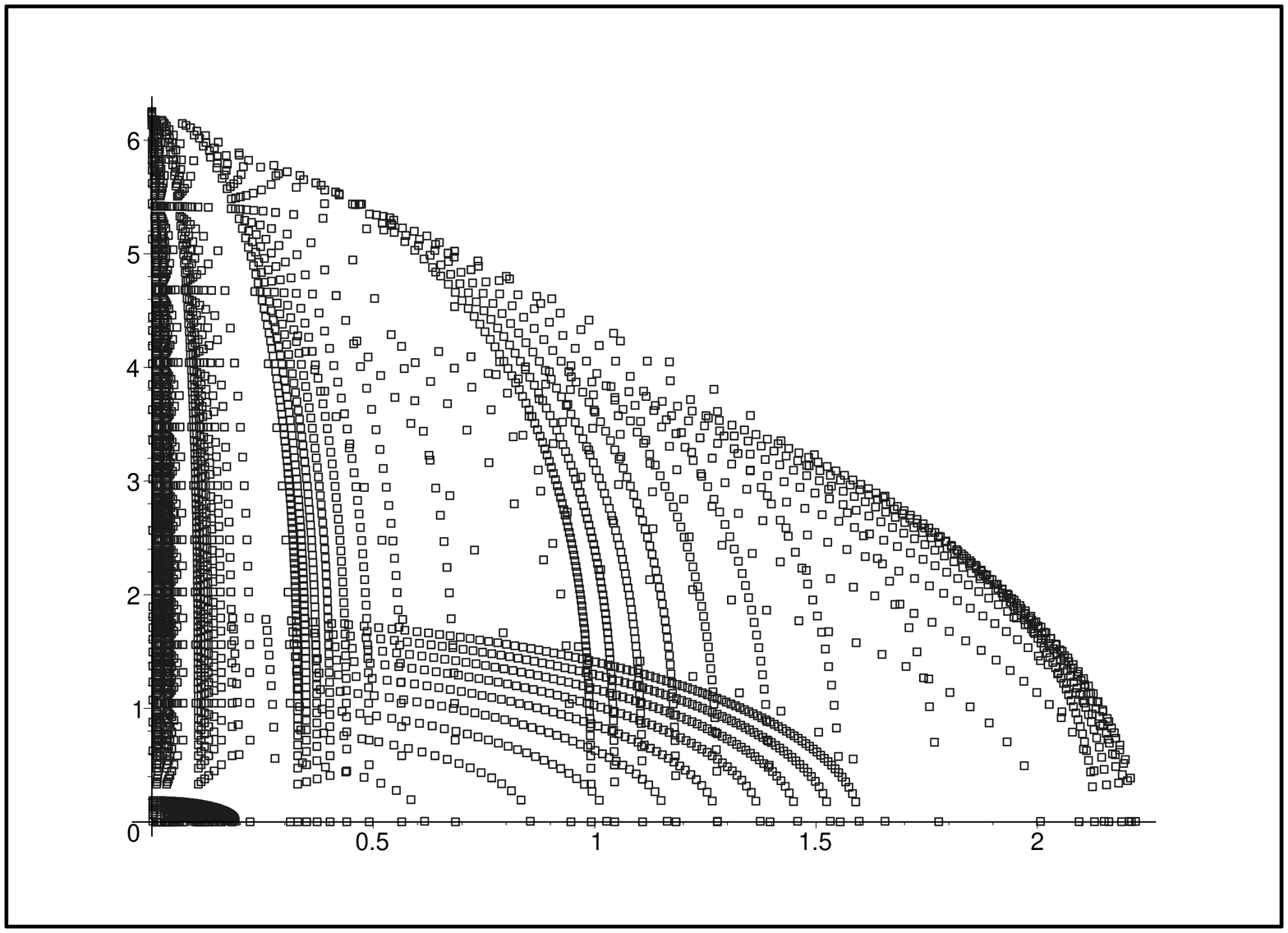,height=3.5in,angle=-90}}
         \vskip0.5cm
         {\footnotesize Figure 2: The existence of bubbles
         in the  $(\beta,\beta|\phi|)$ plane in the one charge
         case.}
\end{figure}

Existence of the instability requires a negative mode\footnote{We
note that due to the analytic continuation compared to the thermal
interpretation, the existence of negative mode corresponds to a
positive Hessian, instead of negative value.}, and in addition we
need to satisfy the condition $I_{bub}>0$ for metastability. We see
in figure 3 that the region of instability always satisfies the
condition for metastability, the transition is always first order.

\begin{figure}[h]
         \beginlabels\refpos 0 0 {}
                     \put 30 -100 {A}
                     \put 115 -160 {B}
                     \put 40 -60 {C}
                     \put 150 -150 {D}
                     \put 150 -80 {E}
                     \put 460 -188 {\beta}
                     \put 250 -40 {\beta|\phi|}
         \endlabels
         \centerline{
         \psfig{figure=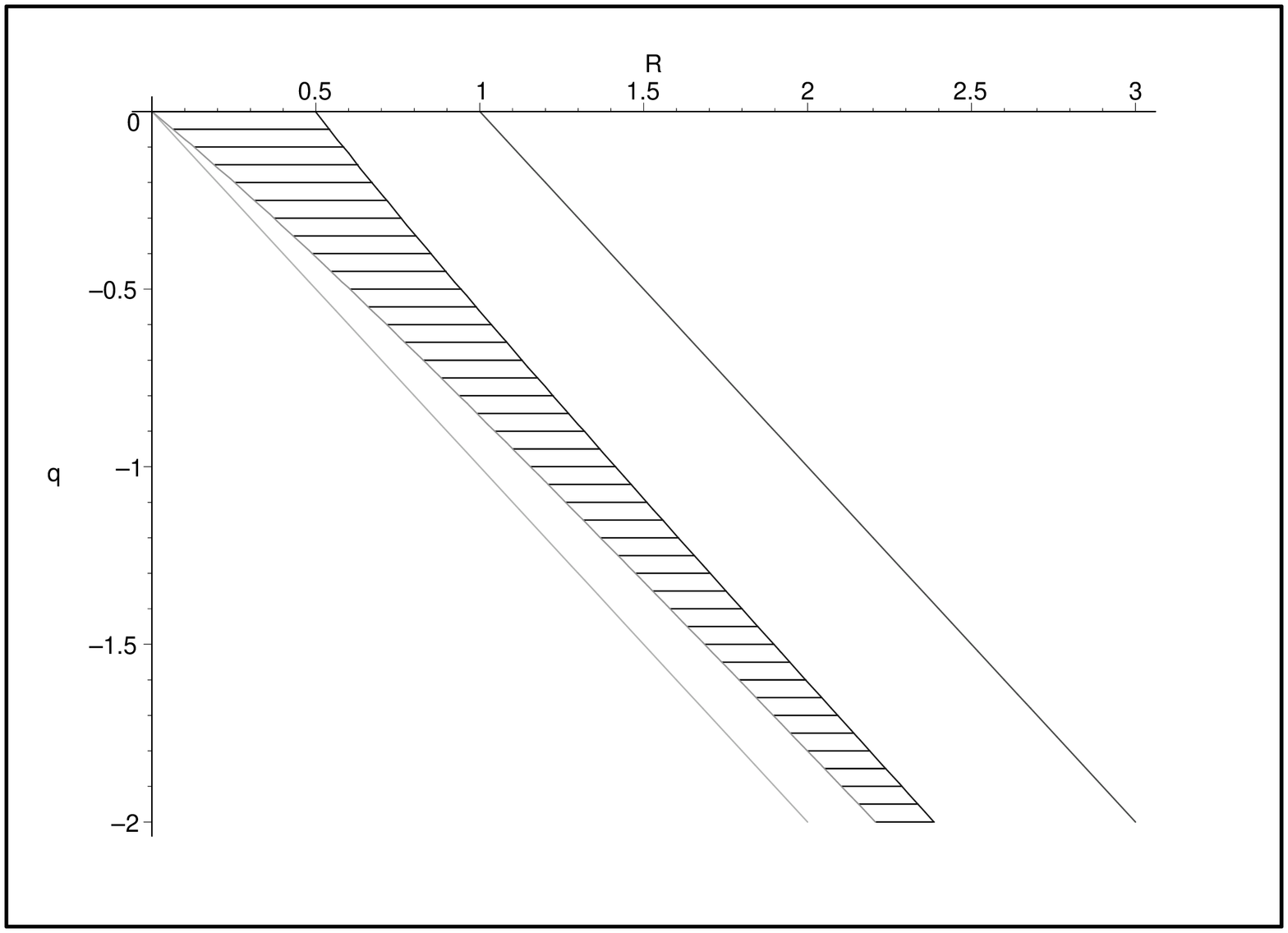,height=3.5in,angle=-90}
         \hskip0.5cm
         \psfig{figure=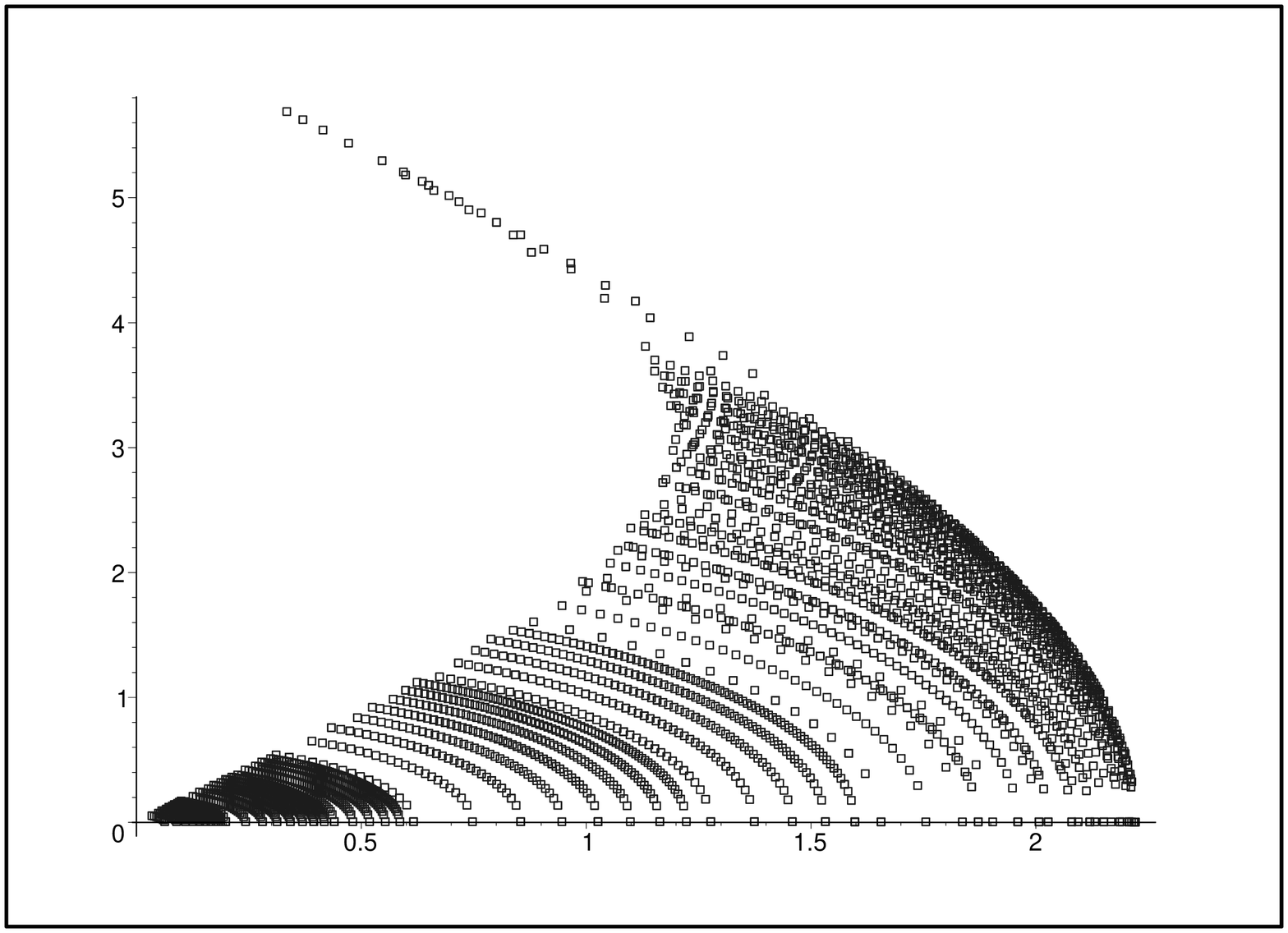,height=3.5in,angle=-90}}
         \vskip0.5cm
         {\footnotesize Figure 3: In the left diagram, instability can exist only in
         region C. No bubbles exist in region A, and region E
         corresponds to disappearance of the barrier $I_{bub}<0$.
          In the right $(\beta,\beta|\phi|)$ diagram, the shaded region describes the region of possible
          instability.}

\end{figure}

Figure 3 shows the existence of instability, and the region of
metastability, in parameter diagrams $(R= r_+^2,q)$ and
$(\beta,\beta|\phi|)$. Interestingly we find both $\beta$ and
$\beta|\phi|$ are bounded. As a check we note that all the features
of the neutral instability are reproduces when setting $\phi=0$.

\subsection{Three Equal Charges}

As an additional example we now consider the case  of three equal
charges, $q_1=q_2=q_3=q$, so that \ba f&=&1-\frac{\mu}{r^2}+r^2 H,
\hspace{1cm}H=(1+\frac{q}{r^2})^3 \ea In this case all the scalars
$X_i$ are constants, therefore the black hole is simply the AdS
Reissner-Nordstrom solution, investigated for example in
\cite{clifford}.

In this case we find that the physical parameters are \ba
\beta&=&\frac{2\pi R(R+q)^{3/2}}{2R^3+R^2(1+3q)-q^3},\nonumber\\
\tilde{q}^2&=&q(R+q)\left(1+\frac{1}{R}(R+q)^2\right),\nonumber\\
\phi&=&-\frac{\tilde{q}}{R+q}. \ea where once again $q<0$,
corresponding to purely imaginary $\phi$, for the bubble
interpretation.

Similar to the discussion above, we first look at the region of
parameters corresponding to the existence of a bubble spacetime.
This is depicted in figure 4.

\begin{figure}[h]
         \beginlabels\refpos 0 0 {}
                     \put 330 -190 {\beta}
                     \put 120 -40 {\beta|\phi|}
         \endlabels
         \centerline{
         \psfig{figure=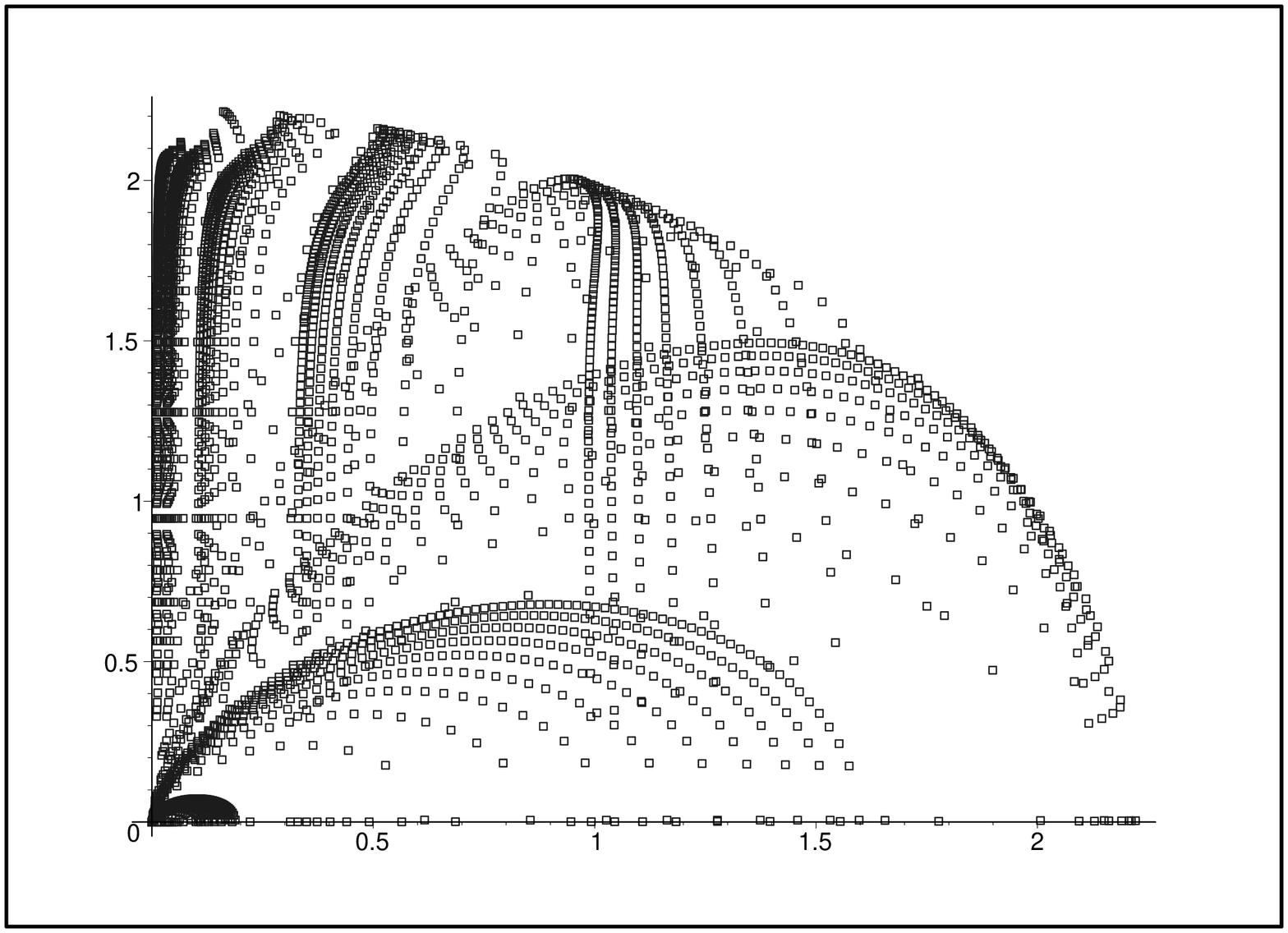,height=3.5in,angle=-90}}
         \vskip0.5cm
         {\footnotesize Figure 4: The existence of bubbles
         in the  $(\beta,\beta|\phi|)$ plane in the three equal charges case.
          }
\end{figure}

Finally, in figure 5 we show the regions for which the negative mode
exists, and the region for which the topological black hole is
meta-stable. As before we see that instability always occurs when
the condition for metastability is satisfied, in other words the
transition is always first order, and proceeds by tunneling over a
potential barrier.

\begin{figure}[h]
         \beginlabels\refpos 0 0 {}
                     \put 70 -140 {A}
                     \put 120 -110 {B}
                     \put 160 -110 {C}
                     \put 160 -70 {D}
                     \put 460 -190 {\beta}
                     \put 250 -40 {\beta|\phi|}
         \endlabels
         \centerline{
         \psfig{figure=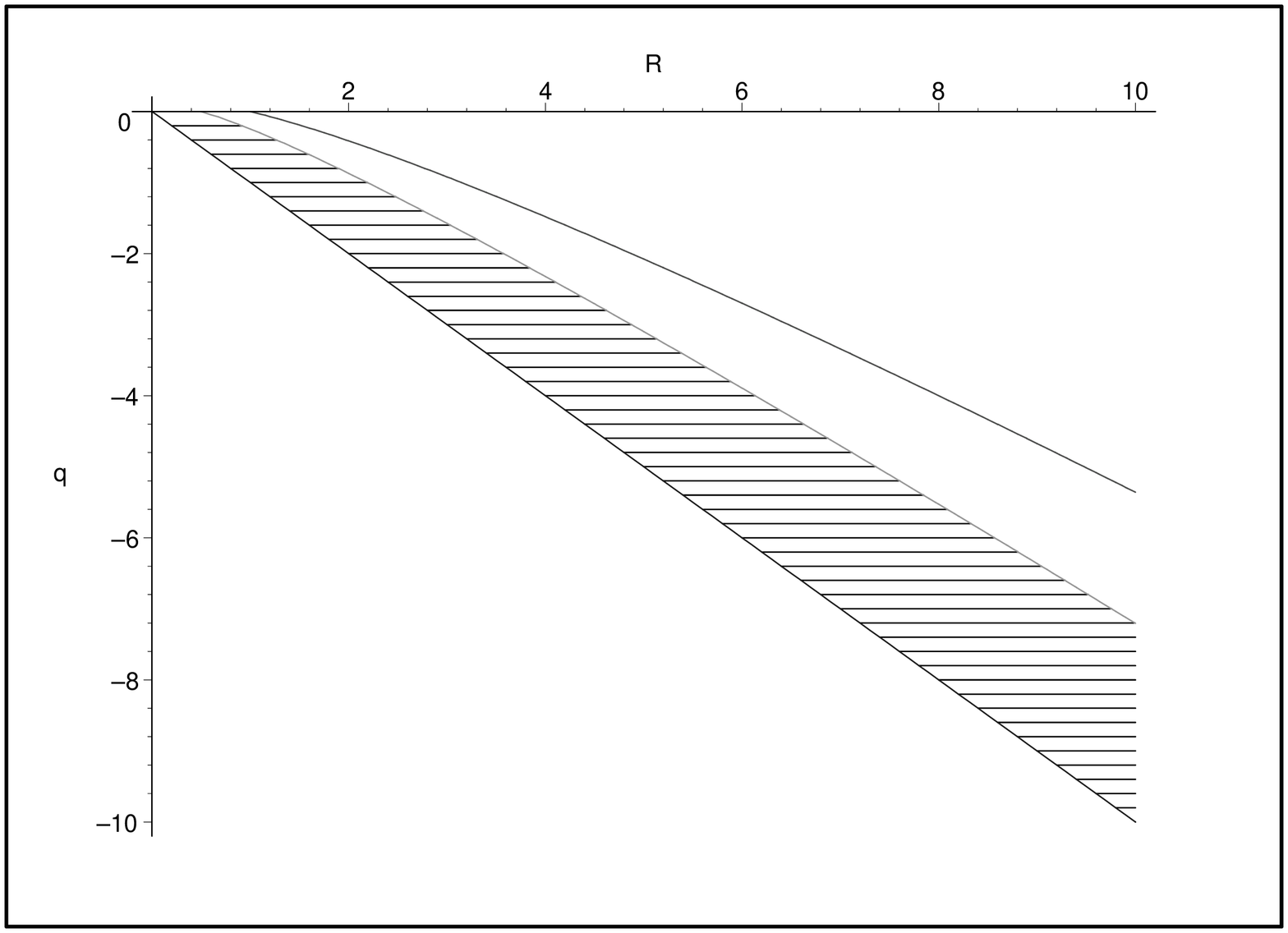,height=3.5in,angle=-90}
         \hskip0.5cm
         \psfig{figure=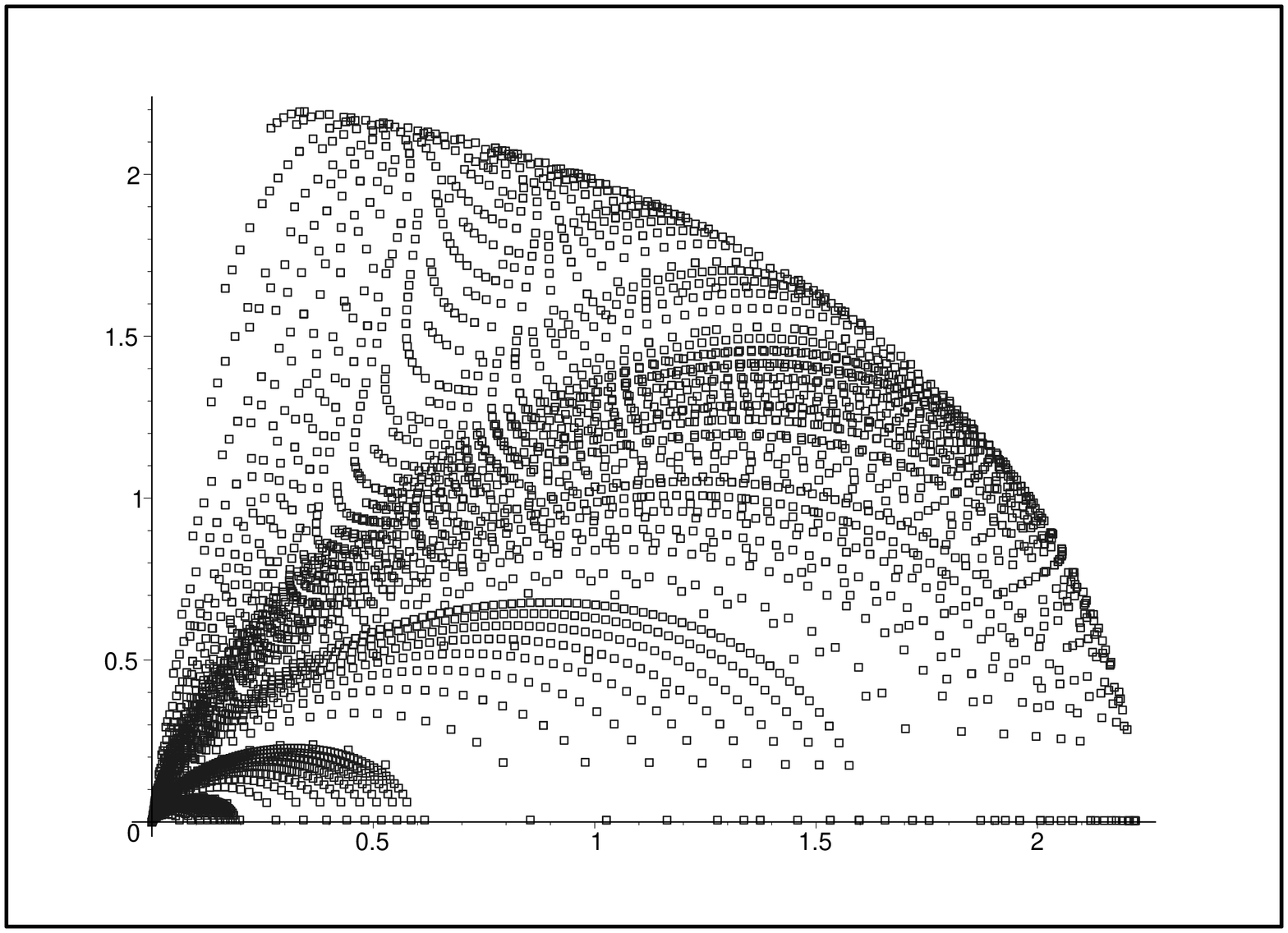,height=3.5in,angle=-90}}
         \vskip0.5cm
{\footnotesize Figure 5: In the left diagram, instability can exist
only in
         region B. No bubbles exist in region A, and region D
         corresponds to disappearance of the barrier $I_{bub}<0$.
          In the right $(\beta,\beta|\phi|)$ diagram, the shaded region describes the region of possible
          instability.}
\end{figure}

\subsection{Features of the Phase Diagrams}

Let us discuss qualitative features of the phase diagram, and how
those are interpreted in terms of the quantum mechanics of a single
variable (the winding condensate) and its effective potential. The
winding mode (Wilson line) $U$ is the lowest lying mode when
compactifying on the circle, thus one can describe the long distance
physics in terms of  three dimensional scalar field theory living on
$dS_3$ space.

We suggest that the physics of bubble nucleation can be effectively
described in terms of the quantum mechanics of the spatially
homogenous mode of $U$, integrating out the massive mode on the
sphere $S^2$ in global coordinates. This approximation will
manifestly break down at late times where the $S^2$ decompactifies,
but may be sufficient for describing the bubble nucleation and early
evolution.

Furthermore, since the bubble is nucleated at $t=0$ when $dS_3$ is
momentarily stationary, we expect that for the process of the bubble
nucleation the time dependence is unimportant\footnote{For
discussions of issues to do with time dependence in the bubble
spacetime see for example \cite{time1,time2}.}, though it may be
important for a complete description of its subsequent evolution. As
further evidence we note that the analysis in \cite{copsey} of the
asymptotic $R_t \times S^2 \times S^1$ boundary suggests that
qualitative features to do with bubble nucleation  are insensitive
to the time dependence. Therefore we suggest to describe the process
with the aid of quantum mechanics with an approximately
time-independent Hamiltonian. That Hamiltonian can be calculated in
the weak coupling limit \cite{us}, here we detail its features that
can be read off from the bulk analysis for strong 'tHooft coupling.

One  distinctive qualitative feature in the phase diagrams is the
point where the Euclidean action becomes small, which we would like
to interpret as the disappearance in the barrier in the
potential\footnote{For a similar statement using a spacetime
analysis \cite{dine1} see \cite{dine2}.}. Indeed, this happens for a
small enough circle radius, and we find in all cases that near that
point (where the circle parametrized by $\chi$ becomes string scale)
a tachyon developed, which is described in the quantum mechanics of
$u$, the winding condensate, as  maximum in the effective potential.
Note that the process of the winding tachyon condensation
corresponds precisely to moving along the $u$ axis. This indicates
also that the true vacuum, even in the regimes where the decay is
non-perturbative, is simply a winding mode condensate. The picture
of the decay by the expanding bubble of nothing is therefore the
conventional one, where a bubble of a true vacuum nucleates inside
the false vacuum, and then exponentially expands outwards. The only
novelty is that the true vacuum is a stringy non-geometrical phase.

The other striking feature of the phase diagrams is the
disappearance of the instability. Indeed, in all cases there is a
critical value of $\beta$, the circle's radius, for which the
instability stops. Equivalently, since the boundary theory is
conformally invariant, there is a minimal radius of curvature of the
space $dS_3$ for which the instability stops. In particular, since
the decay happens at the turning point $t=0$ of $dS_3$, there is a
minimal size of the spatial sphere $S^2$, and below that size the
instability stops.

In conventional quantum mechanics of the homogeneous mode $u$, the
natural explanation of that point is that the false and true vacua
become degenerate in energy, and the instability stops. The bounce
in that case disappears, while its limiting action stays finite.
This is exactly the behavior we find when discussing the bounce near
the critical value of $\beta$.

The quantum mechanics of the homogeneous mode is not sensitive to
spatial dependence of the process. However, using the Euclidean
black hole as an instanton leads to a puzzle related to its deSitter
symmetry. This symmetry  means that the decay is spatially
homogeneous from the boundary viewpoint, for example the expectation
value of the Wilson loop, $\lb U \rb$ is homogeneous. This is
contrary to  the expectation one has from conventional bubble
nucleation in first order transitions\footnote{We thank Don Marolf
for useful comments regarding this point.}, where the typical
situation involves an inhomogeneous pattern in space.

This pattern of decay is likely to be an artifact of the gravity
limit. For bubble nucleation in finite space one has to compare the
size of the space to the radius of a typical bubble. The latter is
determined in terms of parameters of the potential, which are a
function of the 'tHooft coupling $\lambda$. The suppression of the
inhomogeneous decay is then an indication that typical bubble size
is much larger than the size of the space, at least for large enough
$\lambda$. In this case the decay will proceed at once throughout
space.

In order to discover inhomogeneous decays in the gravity dual, we
would have to construct solutions which break the deSitter symmetry
(or the spherical symmetry in Euclidean space), and have lower
action than the spherically symmetric case. However, from bulk
considerations, such instantons are unlikely to dominate at the
gravity limit. The dual field theory interpretation leads us to
conjecture then that at small enough $\lambda$ there ought to be a
transition to non-spherically symmetric bubbles, which will then
dominate the decay process. Such transition is reminiscent of the
Gregory-Laflamme instability, though in this case the localization
is on a sphere.

\section{Conclusions}

We have studied the phase diagram of the maximally supersymmetric
gauge theory on $dS_3 \times S^1$ with various configurations of
gauge fields on the non-contractible circle. The results were
interpreted in terms of quantum mechanics for a single variable
(expectation value of the Wilson line around the circle). We have
seen that various boundaries in those phase diagrams can be
interpreted as features of the effective potential of that quantum
mechanical system. This supports an interpretation of the decay
mediated by the bubble of nothing as conventional vacuum decay, and
the interpretation of the core of the bubble as the true vacuum,
namely the tachyon condensate. This agrees with the interpretation
of the tachyon condensation in the case of the AdS soliton, which is
a limit of the bubble of nothing solutions we discuss.

Even though it is plausible that the dynamics effectively reduces to
that of a single degree of freedom, we are not able to calculate the
effective potential in the strong 'tHooft coupling regime. It is
natural to attempt such calculation in other regimes of the gauge
theory, and compare the qualitative features to those obtained here.
A study of the same gauge theory at weak coupling is currently
underway \cite{us}. In addition to providing a definition of the
effective potential, this study can shed some light on the
mysterious "nothing" state, by viewing it from a conventional
quantum mechanical perspective.

Additionally, it would be fascinating to probe the nothing state
that apparently exists in the bulk bubble spacetime. It is likely
that the picture in \cite{kutasov} applies here as well, and at the
core of the bubble we have, in addition to the geometrical
description, a winding mode condensate. String scattering in the
background will then probe the winding condensate and will give
indication that the geometrical description of the spacetime is
incomplete. We hope to return to these issues in the near future.

\section*{Acknoledgements}

We thank  Vijay Balasubramanian, Micha Berkooz, Michael Dine, Gary
Horowitz, Don Marolf, Rob Myers, Joan Simon,  Evgeny Sorkin and Mark
van Raamsdonk for useful and enjoyable conversations. This work is
supported by discovery grant from NSERC of Canada.

\end{document}